\documentstyle[11pt]{article}

\newcommand{\be}{\begin{displaymath}}
\newcommand{\ee}{\end{displaymath}}
\newcommand{\bea}{\begin{eqnarray}}
\newcommand{\eea}{\end{eqnarray}}
\newcommand{\bean}{\begin{eqnarray*}}
\newcommand{\eean}{\end{eqnarray*}}
\newcommand{\bes}{\begin{subequations}}
\newcommand{\dis}{\displaystyle}
\newcommand{\ees}{\end{subequations}}
\newcommand{\commentout}[1]{}

\newcommand{\pdr}[2]{\frac{\partial{#1}}{\partial{#2}}}

\newcommand{\vmu}{\mbox{\boldmath{$\mu$}}}
\newcommand{\vnu}{\mbox{\boldmath{$\nu$}}}

\newcommand{\vz}{{\bf z}}
\newcommand{\bk}{{\bf k}}

\newcommand{\vx}{{\bf x}}

\newcommand{\vy}{{\bf y}}

\newcommand{\ve}{{\bf e}}

\newcommand{\eps}{{\varepsilon}}

\def\boxit#1{\vbox{\hrule\hbox{\vrule\kern3pt
\vbox{\kern3pt#1\kern3pt}\kern3pt\vrule}\hrule}}

\newcommand{\iint}{\int \!\! \int}

\newcommand{\al}{{\alpha}}
\newcommand{\bq}{{\bf q}}
\newcommand{\bp}{{\bf p}}

\begin{document}

\title{Radiative Transport in a Periodic Structure 
}

\author{Guillaume Bal 
\thanks{Department of Mathematics, Stanford CA, 94305; bal@math.stanford.edu}
\and
Albert Fannjiang
\thanks{Department of Mathematics, University of California Davis, Davis CA, 95616;
fannjian@math.ucdavis.edu}
\and 
George Papanicolaou
\thanks{Department of Mathematics, Stanford CA, 94305; papanico@math.stanford.edu}
\and Leonid Ryzhik
\thanks{Department of Mathematics, University of Chicago, Chicago IL, 60637; 
ryzhik@math.uchicago.edu}
}

\maketitle

\begin{abstract}  
We derive radiative transport equations for solutions of a Schr\"odinger
equation in a periodic structure with small random inhomogeneities.
We use systematically the Wigner transform and the Bloch wave expansion.
The streaming part of the radiative transport equations is determined
entirely by the Bloch spectrum, while the scattering part by the random
fluctuations.

\end{abstract}

\section{Introduction}

Radiative transport equations describe propagation of the phase space
energy density of high frequency waves in a medium with weak random
impurities whose correlation length is comparable to the wave length
$\lambda$ \cite{RPK-WM}.  The background medium may vary but only on
scales that are much larger than $\lambda$. The phase space energy
density has also been studied for a periodic potential when the period
is comparable to the wave length \cite{GMMP,MMP}.  It is then found
that the limiting phase space energy density satisfies a system of
decoupled Liouville equations with Hamiltonians given by the Bloch
eigenvalues.  Thus, we expect that the addition of small random
fluctuations to the periodic structure will give rise to a system of
coupled radiative transport equations. This is what we show in this
paper formally, using asymptotic expansions. The rigorous proof of our
final result remains an open problem. Our main result is the system of
the radiative transport equations (\ref{rte-r}) that describes the
propagation of phase space energy densities $\sigma_m(t,\vx,\bp)$
corresponding to various Bloch eigenvalues $E_m(\bp)$. We derive here
the transport equations only for the Schr\"odinger equation. However,
the generalization to more general types of waves in a periodic and
random medium, like those described by hyperbolic systems considered
in \cite{RPK-WM}, is straightforward.  This  results in the
replacement of the eigenvalues $E_m(\bp)$ in (\ref{rte-r}) by the
corresponding Bloch eigenvalues for the hyperbolic system under
consideration. The paper is organized is follows. First, we recall in
Section 2 some properties of the Bloch eigenfunctions and also give a
formal derivation of the Liouville transport equations in the absence
of random inhomogeneities. These results were previously derived
rigorously in \cite{MMP} and \cite{GMMP}, and we show the connection
between our formalism and their results in Section 2.4. Section 3 is
the main part of the paper, where we derive the radiative transport
equations in the presence of a random potential.

\section{Waves in a periodic structure}

\subsection{The Schr\"odinger equation}

We give here a derivation of the Liouville equation for the phase space
energy density like the one in \cite{RPK-WM}. 
A different analysis is given in \cite{GMMP,MMP}.
We then adapt the analysis of \cite{RPK-WM} to the periodic-random case.

There is not a lot of mathematical work on the transport limit for the
Schr\"odinger equation with random potential. We cite here the work of
Martin and Emch \cite{ME}, of Spohn \cite{Spohn}, of Dell'Antonio
\cite{DA} and the recent extensive study of Ho, Landau and Wilkins
\cite{HLW}. These papers established the validity of the kinetic
linear transport equation for a small time $T>0$.  The global validity
of this limit was proved recently by Erd\"os and Yau in
\cite{Erdos-Yau}. They treat only spatially homogeneous problems but
it is known how to extend the analysis to the spatially inhomogeneous
case (slow $\vx$-dependent initial data and potential) \cite{AF}. A
really satisfactory mathematical treatment of radiative transport
asymptotics from random wave equations is lacking at present.

It is convenient for us to 
use the usual Wigner distribution and not the Wigner band-series
as in \cite{GMMP,MMP}. The 
two formulations are, however, equivalent \cite{MMP}.
Let $\phi_\eps(t,\vx)$ be the 
solution of the initial value problem
\bea\label{schrod}
i\eps\pdr{\phi_\eps}{t}
+\frac{\eps^2}{2}\Delta\phi_\eps-V(\frac{\vx}{\eps})&=&0\\
\phi_\eps(0,\vx)&=&\phi_\eps^0(\vx).\nonumber
\eea
The initial data $\phi_\eps^0(\vx)$ is uniformly bounded in $L^2(R^d)$: 
$\dis\|\phi_\eps^0\|_{L^2}\le C$. We also assume that it is
$\eps$-oscillatory, that is, for any test function $\psi\in C_c(R^d)$:
\bea\label{epsoscill}
\limsup_{\eps\to 0}\int_{|\bk|\ge R/\eps}|\widehat{\psi\phi_\eps^0}(\bk)|^2d\bk
\to 0\hbox{   as $R$ goes to $+\infty$}.
\eea
Finally we assume that the family $\phi_\eps^0$ is compact at infinity:
\bea\label{comp-infty}
\limsup_{\eps\to 0}\int_{|\vx|\ge R}|\phi_\eps^0(\vx)|^2d\vx
\to 0\hbox{   as $R$ goes to $+\infty$}.
\eea
A sufficient condition for (\ref{epsoscill}) is 
$\dis\|\nabla\phi_\eps^0||\le C/\eps$. 
The case of $\eps$-independent initial data was studied in \cite{BLP}, where
the Liouville equation for the wave amplitudes (not energies) was derived.

The potential $V(\vz)$ is periodic:
\be
V(\vz+\vnu)=V(\vz),
\ee
where vector $\vnu$ belongs to the period lattice $L$:
\bea\label{lattice}
L=\left\{\sum_{j=1}^d n_j\ve_j|~n_j\in Z\right\},
\eea
and $\ve_1,\dots,\ve_d$ form a basis of $R^d$ with the dual basis
$\ve^j$ defined by
\be
(\ve_j\cdot\ve^k)=2\pi\delta_{jk} 
\ee
and the dual lattice $L^*$ defined by (\ref{lattice}) with $\ve_j$ replaced
by $\ve^j$. We denote by $C$ the basic period cell of $L$ and by $B$
the Brillouin zone:
\be
B=\left\{\bk\in R^d|~\bk \hbox{ is closer to 
\vmu=0 than any other point $\vmu\in L^*$}
\right\}.
\ee
We define the Wigner distribution by
\bea\label{wigner-00}
W_\eps(t,\vx,\bk)=\int_{R^d}\frac{d\vy}{(2\pi)^d}e^{i\bk\cdot\vy}
\phi_\eps(t,\vx-{\eps\vy})\bar\phi_\eps(t,\vx).
\eea
This definition is equivalent to its symmetric version
\[
\tilde W_\eps(t,\vx,\bk)=\int_{R^d}\frac{d\vy}{(2\pi)^d}e^{i\bk\cdot\vy}
\phi_\eps(t,\vx-\frac{\eps\vy}{2})\bar\phi_\eps(t,\vx+\frac{\eps\vy}{2}),
\]
in the sense that $W_\eps$ and $\tilde W_\eps(\vx,\bk)$ have the same weak 
limit as $\eps\to 0$ \cite{GL}. We also have that
\[
{\cal E}_\eps(t,\vx)=|\phi_\eps(t,\vx)|^2=\int_{R^d}d\bk W_\eps(t,\vx,\bk)=
\int_{R^d}d\bk \tilde W_\eps(t,\vx,\bk).
\]
The basic properties of the Wigner distributions are reviewed in
detail in \cite{GMMP,CPDE}. In particular the weak limit
$W(t,\vx,\bk)$ of $W_\eps(t,\vx,\bk)$ exists in ${\cal S}'(R^d\times
R^d)$ and under assumptions (\ref{epsoscill}) and (\ref{comp-infty}) it
captures correctly the behavior of the energy ${\cal E}_\eps$:
\[
\lim_{\eps\to 0}\int d\vx{\cal E}_\eps(t,\vx)=\iint d\vx d\bk W(t,\vx,\bk).
\]

 We deduce from (\ref{schrod}) and (\ref{wigner-00})
 the following evolution equation for 
$W_\eps(t,\vx,\bk)$:
\bea\label{wig-eq}
\pdr{W_\eps}{t}+\bk\cdot\nabla_{\vx}W_\eps+\frac{i\eps}{2}\Delta_{\vx}W_\eps
=\frac{1}{i\eps}
\sum_{\vmu\in L^*}e^{i\vmu\cdot\vx/\eps}
\hat{V}(\vmu)\left[W_\eps(\vx,\bk-{\vmu})-
W_\eps(\vx,\bk)\right].
\eea
Here $\hat{V}(\vmu)$ are the periodic Fourier coefficients of $V(\vy)$:
\bea\label{f-per}
\hat{V}(\vmu)=\frac{1}{|C|}\int_Cd\vy e^{-i\vmu\cdot\vy}V(\vy),
\eea
so that if $V(\vy)$ is smooth we have
\be
V(\vy)=\sum_{\vmu\in L^*}e^{i\vmu\cdot\vy}\hat{V}(\vmu)
\ee
and
\bea\label{delta-mu}
\frac{1}{|C|}\sum_{\mu\in L^*}e^{i\vmu\cdot\vz}=
\sum_{\vnu\in L}\delta(\vz-\vnu).
\eea
We introduce a multiple scales expansion for $W_\eps$:
\bea\label{pert-exp}
W_\eps(t,\vx,\bk)=W_0(t,\vx,\frac{\vx}{\eps},\bk)+
\eps W_1(t,\vx,\frac{\vx}{\eps},\bk)+\dots
\eea
and assume that the leading term $W_0(t,\vx,\vz,\bk)$ 
is periodic in the fast variable 
$\dis\vz=\frac{\vx}{\eps}$. As usual, we replace then
\be
\nabla_{\vx}\to\nabla_{\vx}+\frac{1}{\eps}\nabla_{\vz}
\ee
in (\ref{wig-eq}) and rewrite it as 
\bea\label{wig-eq-2}
\pdr{W_\eps}{t}&&+\bk\cdot\left[\nabla_{\vx}+\frac{1}{\eps}\nabla_\vz\right]
W_\eps+
\frac{i\eps}{2}(\nabla_\vx+\frac{1}{\eps}\nabla_\vz)\cdot
(\nabla_\vx+\frac{1}{\eps}\nabla_\vz)W_\eps\\&&=\frac{1}{i\eps}
\sum_{\vmu\in L^*}e^{i\vmu\cdot\vz}
\hat{V}(\vmu)\left[W_\eps(\bk-{\vmu})-
W_\eps(\bk)\right].~~~~~~\nonumber
\eea
We insert the perturbation expansion (\ref{pert-exp}) into (\ref{wig-eq-2})
and get at the order $\eps^{-1}$:
\bea\label{order1}
{\cal L} W_0=0,
\eea
where  the skew symmetric operator $\cal L$ is given by
\be
{\cal L}f(\vz,\bk)=\bk\cdot\nabla_{\vz}f+\frac{i}{2}\Delta_\vz f-\frac{1}{i}
\sum_{\vmu\in L^*}e^{i\vmu\cdot\vz}
\hat{V}(\vmu)\left[f(\vz,\bk-{\vmu})-
f(\vz,\bk)\right].
\ee

\subsection{The Bloch functions}

The eigenfunctions of $\cal L$ are constructed as follows. Given $\bp\in R^d$
consider the eigenvalue
problem 
\bea\label{bloch}
-\frac 12\Delta_{\vz}\Psi(\vz,\bp)+V(\vz)\Psi(\vz,\bp)&=&E(\bp)\Psi(\vz,\bp)\\
\Psi(\vz+\vnu,\bp)&=&e^{i\bp\cdot\vnu}\Psi(\vz,\bp),~\hbox{for all }
\vnu\in L\nonumber\\
\pdr{\Psi}{z_j}(\vz+\vnu,\bp)&=&
e^{i\bp\cdot\vnu}\pdr{\Psi}{z_j}(\vz),~\hbox{for all }\vnu\in L.
\nonumber
\eea
This problem has a complete orthonormal
 basis of eigenfunctions $\Psi_m^\al(\vz,\bp)$ in $L^2(C)$:
\bea\label{orth-psi}
(\Psi_m^\al,\Psi_j^\beta)=
\int_C\frac{d\vz}{|C|}\Psi_m^\al(\vz,\bp)\bar\Psi_j^\beta(\vz,\bp)=\delta_{mj}
\delta_{\al\beta}.
\eea
They are called the Bloch eigenfunctions,
corresponding to the real 
eigenvalues $E_m(\bp)$ of multiplicity $r_m$. Here $\al=1,\dots,r_m$ labels
eigenfunctions inside the eigenspace. The eigenvalues $E_m(\bp)$ are 
$L^*$-periodic in $\bp$ and have constant
finite multiplicity outside a closed subset $F_m$ of $\bp\in R^d$ 
of measure zero. They may be arranged
$E_1(\bp)< E_2(\bp)<\dots < E_j(\bp)<\dots$ with $E_j(\bp)\to\infty$
as $j\to\infty$, uniformly in $\bp$ \cite{BLP,Keller-Odeh,WX}. We consider
momenta $\bp$ outside the set $F_m$.

The problem (\ref{bloch}) may be rewritten in terms of periodic functions
$\Phi(\vz,\bp)=e^{-i\bp\cdot\vz}\Psi(\vz,\bp)$:
\bea\label{bloch2}
-\frac 12\Delta_{\vz}\Phi_m^\al+V(\vz)\Phi_m^\al
-i\bp\cdot\nabla_{\vz}\Phi_m^\al+
\frac{|\bp|^2}{2}\Phi_m^\al=E_m(\bp)\Phi_m^\al(\vz,\bp).
\eea
We differentiate (\ref{bloch2}) with respect to $p_j$ and take the  scalar
product of the resulting equation and $\Phi_m^\beta$ to get
\be
\pdr{E_m}{p_j}\delta_{\al\beta}=p_j \delta_{\al\beta}
-i(\pdr{\Phi_m^\al}{z_j},\Phi_m^\beta),
\ee
which may be rewritten as 
\bea\label{der-of-em}
\pdr{E_m}{p_j}\delta_{\al\beta}=i(\Psi_m^\al,\pdr{\Psi_m^\beta}{z_j}).
\eea
There is no summation over $m$ in (\ref{der-of-em}).

Recall the Bloch transform of a function 
$\phi(x)\in L^2(R^d)$ 
\[
\tilde \phi_m^\al(\bp)=\int_{R^d}d\vz\phi(\vz)\bar\Psi_m^\al(\vz,\bp).
\]
It has the following properties:
\begin{itemize}
\item[(i)]{$\displaystyle\phi(\vx)=\frac{1}{|B|}\sum_{m=1}^\infty
\sum_{\al=1}^{r_m}
\int_Bd\bp\tilde\phi_m^\al(\bp)\Psi_m^\al(\vx,\bp),~~\vx\in R^d$.}
\item[(ii)]{ Let $\phi(\vx),\eta(\vx)\in L^2(R^d)$, then the
Plancherel formula holds:
\be
\int_{R^d}d\vx\phi(\vx)\bar\eta(\vx)=\frac{1}{|B|}\sum_{m,\al}\int_B
d\bp\tilde\phi_m^\al(\bp)\overline{\tilde\eta_m^\al(\bp)}.
\ee
}
\item[(iii)]{The mapping $\phi\to\tilde\phi$ is one-to-one and onto, from
$L^2(R^d)\to\oplus_mL^2(B)$.}
\end{itemize}
We deduce from these properties the orthogonality relations:
\be
\delta(\vy-\vx)=\frac{1}{|B|}\sum_{m,\al}
\int_Bd\bp\Psi_m^\al(\vx,\bp)\bar\Psi_m^\al(\vy,\bp)
\ee
and
\bea\label{p-ort}
\delta_{jm}\delta_{\al\beta}\delta_{per}(\bp-\bq)=
\frac{1}{|B|}\int_{R^d}d\vx\Psi_j^\al(\vx,\bp)\bar\Psi_m^\beta(\vx,\bq).
\eea
The periodic delta function $\delta_{per}$ in (\ref{p-ort}) is understood
as follows: for any function $\phi(\bp)\in C^\infty(B)$
\[
\phi(\bp)=\int_Bd\bq\phi(\bq)\delta_{per}(\bp-\bq).
\]

Given any vector $\bk\in R^d$ we may decompose it uniquely as 
\begin{equation}\label{kmup}
\bk=\bp_\bk+\vmu_\bk
\end{equation}
with $\bp_\bk\in B$ and $\vmu_\bk\in L^*$.
We then define the $\vz$-periodic functions $Q_{mn}^{\al\beta}(\vz,\vmu,\bp)$, 
$\vmu\in L^*$, $\bp\in B$ by
\bea\label{qmn}
Q_{mn}^{\alpha\beta}(\vz,\vmu,\bp)=
\int_C\frac{d\vy}{|C|}e^{i(\bp+\vmu)\cdot\vy}
\Psi_m^{\al}(\vz-{\vy},\bp)\bar\Psi_n^\beta(\vz,\bp).
\eea
Then a direct computation
shows that
\bea\label{l-on-qmn}
{\cal L}Q_{mn}^{\al\beta}(\vz,\vmu,\bp)=
i(E_m(\bp)-E_n(\bp))
Q_{mn}^{\al\beta}(\vz,\vmu,\bp)
\eea
with $\vmu=\vmu_\bk$, $\bp=\bp_\bk$,
so $Q_{mn}^{\al\beta}$ are eigenfunctions of $\cal L$.

\subsection{The Liouville equations}

Now (\ref{l-on-qmn}) implies that, for any $\bp$, 
$\hbox{ker}{\cal L}$ is spanned
by the functions $Q_{mm}^{\al\beta}$, which we denote by $Q_m^{\al\beta}$
(to indicate that there is no summation over $m$).
Then (\ref{order1}) implies that $W_0(t,\vx,\vz,\bk)$ may be written 
as
\bea\label{sigma}
W_0(t,\vx,\vz,\bk)=
W_0(t,\vx,\vz,\bp+\vmu)=\sum_{m,\al,\beta} \sigma_m^{\al\beta}(t,\vx,\bp)
Q_{m}^{\al\beta}(\vz,\vmu,\bp),~~\bp\in B,~\vmu\in L^*
\eea
with $\vmu=\vmu_\bk$, $\bp=\bp_\bk$.
This defines $\sigma_m$, which is scalar if the eigenvalue $E_m(\bp)$ 
is simple, and is a matrix of size $r_m\times r_m$
 if $E_m(\bp)$ has multiplicity
$r_m>1$. We call $\sigma_m$ the coherence matrices in analogy to the 
non-periodic case \cite{RPK-WM}. They are defined inside the Brillouin zone
$\bp\in B$ but it is convenient to extend them 
as functions in $R^d$, $L^*$-periodic in $\bp$.

Next we look at $\eps^0$ terms in (\ref{wig-eq-2}). We get 
\bea\label{order0}
\pdr{W_0}{t}+\bk\cdot\nabla_{\vx}W_0+i\nabla_{\vx}\cdot\nabla_{\vz}W_0
=-{\cal L}W_1.
\eea
We now integrate both sides of (\ref{order0}) against 
$\bar Q_{j}^{\al\beta}(\vz,\vmu,\bp)$ over $\vz$ in $C$ and sum over $\vmu$.
To evaluate the left side we note that we
have after summing over $\vmu$ using (\ref{delta-mu}), 
and using (\ref{orth-psi}) 
and a change of variables
$\vy\to \vz-{\vy}$:
\bea
\sum_{\vmu\in L^*}
\int_C \frac{d\vz}{|C|}
 Q_m^{\al'\beta'}(\vz,\vmu,\bp)\bar Q_{j}^{\al\beta}(\vz,\vmu,\bp)&=&
\int_{C\times C}
 \frac{d\vy d\vz}{|C|^2}\Psi_m^{\al'}(\vz,\bp)\bar\Psi_m^{\beta'}
(\vy,\bp)\bar\Psi_j^\al(\vz,\bp)\Psi_j^\beta(\vy,\bp)
\nonumber\\
\label{orthogon-q}
&=&\delta_{mj}\delta_{\al\al'}\delta_{\beta\beta'}.
\eea
Further, using the change of variables and equations 
above and also (\ref{der-of-em}), we get
\bea
&&\sum_{\vmu\in L^*}((p_l+\mu_l)Q_m^{\al'\beta'},Q_{j}^{\al\beta})\nonumber\\
&&=
\sum_{\vmu\in L^*}\int_{C^3} \frac{d\vz d\vy_1 d\vy_2}{|C|^3}
e^{i(\bp+\vmu)\cdot(\vy_1-\vy_2)}(p_l+\mu_l)
\Psi_m^{\al'}(\vz-{\vy_1},\bp)\bar\Psi_m^{\beta'}(\vz,\bp) 
\bar\Psi_j^{\al}(\vz-{\vy_2},\bp)
\Psi_j^{\beta}(\vz,\bp)\nonumber\\
&&=
\frac{1}{i}\int_{C\times C}
\frac{d\vz d\vy}{|C|^2}\pdr{\Psi_m^{\al'}(\vy,\bp)}{y_l}
\bar\Psi_j^{\al}(\vy,\bp)
\Psi_j^{\beta}(\vz,\bp)
\bar\Psi_m^{\beta'}(\vz,\bp)\nonumber\\
\label{k-q}&&=
\frac{1}{i}\delta_{mj}\delta_{\beta\beta'}
(\pdr{\Psi_m^{\al'}}{y_l},\Psi_j^{\al})
=\delta_{mj}\delta_{\al\al'}\delta_{\beta\beta'}
\pdr{E_m}{p_l}.
\eea
A similar calculation shows that the third term on the left 
vanishes:
\bea
\sum_{\vmu\in L^*}(\pdr{Q_{m}^{\al'\beta'}}{z_l},Q_{j}^{\al\beta})&&
=\int_{C\times C}\frac{d\vz d\vy}{|C|^2}\pdr{\Psi_m^{\al'}(\vz-\vy,\bp)}{z_l}
\bar\Psi_{j}^{\al}(\vz-\vy,\bp)\Psi_j^{\beta}(\vx,\bp)
\bar\Psi_m^{\beta'}(\vz,\bp)\nonumber\\
&&+\int_{C\times C}\frac{d\vz d\vy}{|C|^2}{\Psi_m^{\al'}(\vz-\vy,\bp)}
\bar\Psi_{j}^{\al}(\vz-\vy,\bp)\Psi_j^{\beta}(\vx,\bp)
\pdr{\bar\Psi_m^{\beta'}(\vz,\bp)}{z_l}\nonumber\\&&
=-\frac{1}{i}\delta_{\al\beta}\delta_{\beta\beta'}\delta_{jm}\pdr{E_m}{p_l}
+\frac{1}{i}\delta_{\al\beta}\delta_{\beta\beta'}\delta_{jm}\pdr{E_m}{p_l}=0.
\nonumber
\eea
The right side of (\ref{order0}) integrated against 
$\bar Q_{j}^{\al\beta}$
vanishes  
since $\cal L$ is skew symmetric, 
and $Q_j^{\al\beta}\in\hbox{ker}{\cal L}$. 
Putting together (\ref{order0}), (\ref{orthogon-q}) and (\ref{k-q})
we obtain the Liouville equations for the coherence matrices $\sigma_m$:
\bea\label{liouville}
\pdr{\sigma_m}{t}+\nabla_{\bp}E_m\cdot\nabla_{\vx}\sigma_m=0.
\eea
The initial data for equations (\ref{liouville}) is constructed as follows.
Let $W_\eps^0(\vx,\bk)$ be the Wigner transform of the initial data $\phi_\eps^0(\vx)$. Then $\sigma_m(0,\vx,\bp)$ is given by
\[
\sigma_m^{\al\beta}(0,\vx,\bp)=\lim_{\eps\to 0}
\sum_{\vmu\in L^*}W_\eps^0(\vx,\bp+\vmu)
\bar Q_j^{\al\beta}(\frac{\vx}{\eps},\vmu,\bp)
\]
with the limit understood in the weak sense.

\subsection{The Wigner band series}

The Liouville equations (\ref{liouville}) were previously derived in 
\cite{MMP,GMMP} in terms of the Wigner series defined by
\[
w_\eps(t,\vx,\bk)=
\frac{|C|}{(2\pi)^d}
\sum_{\vnu\in L}e^{i\bk\cdot\vnu}\phi_\eps(\vx-\eps\vnu)
\bar\phi_\eps(\vx)=\sum_{\vmu\in L^*}W_\eps(\vx,\bk+\vmu).
\]
It was shown that the weak limit $w(t,\vx,\bk)$ of $w_\eps(t,\vx,\bk)$ has 
the form
\[
w(t,\vx,\bk)=\sum_jw_j(t,\vx,\bk).
\]
Here $w_j$ is the limit Wigner series of the projection $\phi_\eps^j$
of the solution $\phi_\eps$ of the Schr\"odinger equation on the 
Bloch spaces ${\cal S}_j^\eps$:
\[
{\cal S}_j^\eps=\left\{f\in L^2(R^d):~~f(\vx)=\sum_{\al}
{1\over |B|}\int_B \frac{d\bp}{\eps^{3/2}}
 \tilde f_\al(\bp)\Psi_j^\al(\frac{\vx}{\eps},\bp)\right\}.
\]
 Each projection $w_j$ evolves according to the Liouville
equation (\ref{liouville}). This result may be related to our approach
as follows. The Wigner distribution $W_\eps(t,\vx,\bk)$ has the asymptotics
\bea\label{withoscill}
W_\eps(t,\vx,\bp+\vmu)\approx\sum_{m,\al,\beta}\sigma_m^{\al\beta}(t,\vx,\bp)
Q_{m}^{\al\beta}
\left(\frac{\vx}{\eps},\vmu,\bp\right).
\eea
Then the weak limit of $w_\eps$ is given by
\[
w(t,\vx,\bp)=\sum_{m,\al,\beta}
\sigma_m^{\al\beta}(t,\vx,\bp)\sum_{\vmu\in L^*}\int_C \frac{d\vz}{|C|}
Q_m^{\al\beta}(\vz,\vmu,\bp)=\sum_{m}
\hbox{Tr}\sigma_m(t,\vx,\bp).
\]
If we take the trace of (\ref{liouville}) we get the transport
equations for $w_m(t,\vx,\bp)=\hbox{Tr}\sigma_m(t,\vx,\bp)$ obtained
in \cite{MMP} and \cite{GMMP}. However, the representation
(\ref{withoscill}) captures not only the weak limit of $w_\eps$ but
also the oscillations of $W_\eps$ on the fine scale. Moreover, the
cross-polarization of various modes corresponding to the same
eigenvalue is also taken into account by the off-diagonal terms in the
coherence matrices. The oscillations of the spatial energy density,
which is the physically interesting quantity, are given by
\[
{\cal E}_\eps(t,\vx,\bp)\approx\sum_{m,\al,\beta}
\int_B d\bp\sigma_m^{\al\beta}(t,\vx,\bp)
\Psi_m^\al(\frac{\vx}{\eps},\bp)\bar\Psi_m^\beta(\frac{\vx}{\eps},\bp)
\rightharpoonup\sum_{m}\int_B d\bp \hbox{Tr}
\sigma_m(t,\vx,\bp)\hbox{~~~as $\eps\to 0$~.}
\]
This information on the energy oscillations may be useful in numerical 
simulations. We see that $\sigma_m^{\al\al}$ are phase space resolved 
energy densities of different modes inside the Brillouin zone.

\section{The random perturbation}

\subsection{Simple eigenvalues }

We assume first that the Bloch eigenvalues $E_m(\bp)$ have
multiplicity one for all $\bp\in B$. This assumption is known to be
true for the leading eigenvalue when the Fourier transform (\ref{f-per})
of the periodic potential $V(\vy)$ is negative \cite{AS}. In many
physical problems the absence of level crossings restricts our results
to the lower part of the spectrum.
We consider now small random perturbations of the periodic problem 
(\ref{schrod}) with randomness being on the same scale as the periodic
potential but weak:
\bea
\label{schro-rand}
i\eps\pdr{\phi_\eps}{t}
+\frac{\eps^2}{2}\Delta\phi_\eps-V(\frac{\vx}{\eps})-\sqrt{\eps}
N(\frac{\vx}{\eps})=0&& \nonumber\\
\phi_\eps(0,\vx)=\phi_\eps^0(\vx)&&.\nonumber
\eea
Here $N(\vy)$ is a time independent mean zero 
spatially homogeneous random process
with covariance tensor $R(\vx)$ defined by:
\bea\label{covar-r}
\langle N(\vy)N(\vy+\vx)\rangle=R(\vx),~~
\langle\hat N(\bp)\hat N(\bq)\rangle=(2\pi)^d\hat R(\bq)\delta(\bp+\bq).
\eea
Here $\langle\cdot\rangle$ denotes the ensemble average and the Fourier
transform $\hat N(\bq)$ is
\bea\label{f-whole}
\hat N(\bq)=\int_{R^d}d\vx e^{-i\bq\cdot\vx}N(\vx).
\eea

The Wigner distribution $W_\eps(t,\vx,\bk)$ satisfies the evolution
equation
\bea
\nonumber
\pdr{W_\eps}{t}+\bk\cdot\nabla_{\vx}W_\eps+\frac{i\eps}{2}\Delta_{\vx}W_\eps&=&
\frac{1}{i\eps}\sum_{\vmu\in L^*}e^{i\vmu\cdot\vx/\eps}\hat{V}(\vmu)\left[
W_\eps(\bk-\vmu)-W_\eps(\bk)\right]\\
&+&\frac{1}{i\sqrt{\eps}}\int_{R^d}\frac{d\bq}{(2\pi)^d}e^{i\bq\cdot\vx/\eps}
\hat N(\bq)\left[W_\eps(\bk-\bq)-W_\eps(\bk)\right].
\nonumber
\eea
Here $\hat V(\vmu)$ is the periodic Fourier transform (\ref{f-per}) and 
$\hat N(\bq)$ is the Fourier transform (\ref{f-whole}) over $R^d$.
We consider the asymptotic expansion
\be
W_\eps(t,\vx,\bk)=W_0(t,\vx,\frac{\vx}{\eps},\bk)+\sqrt{\eps}
W_1(t,\vx,\frac{\vx}{\eps},\bk)+\eps W_2(t,\vx,\frac{\vx}{\eps},\bk)
+\dots,
\ee
with the leading order term $W_0$ being deterministic. We 
introduce as before the fast variable $\vz=\dis\frac{\vx}{\eps}$,
replace $\dis\nabla_\vx\to\nabla_\vx+\frac{1}{\eps}\nabla_{\vz}$
and collect the powers of $\eps$. The order $\eps^{-1}$ gives as before
\be
{\cal L}W_0=0.
\ee
Thus we still have the decomposition (\ref{sigma}): 
\bea\label{decomp-3}
W_0(t,\vx,\vz,\bp+\vmu)=\sum_m\sigma_m(t,\vx,\bp)Q_m(\vz,\vmu,\bp)
\eea
with $\sigma_m$ being scalar because the spectrum is simple. 
Recall that the functions $\sigma_m(t,\vx,\bp)$ are $L^*$-periodic in $\bp$
and the $\vz$-periodic functions $Q_m$ are given by (\ref{qmn}) with $m=n$.
The order $\eps^{-1/2}$ terms give
\bea\label{order1/2-r}
{\cal L}W_1=\frac{1}{i}\int_{R^d}\frac{d\bq}{(2\pi)^d}e^{i\bq\cdot\vz}
\hat N(\bq)\left[W_0(\vz,\bk-{\bq})-W_0(\vz,\bk)
\right].
\eea
The distribution $W_1$ need not be periodic in the fast variable $\vz$.
Therefore it may not be expanded in $Q_{mn}$ and we  
use the  the basis functions
\bea\label{def-pmn}
P_{mn}(\vz,\vmu,\bp,\bq)=\int_C\frac{d\vy}{|C|}e^{i(\bp+\vmu)\cdot\vy}
\Psi_m(\vz-\vy,\bp)\bar\Psi_m(\vz,\bp+\bq),
\eea
defined for $\vz\in R^d$ and 
$\bp,~\bq\in B$, in place of the periodic functions
$Q_{mn}(\vz,\vmu,\bp)$. The functions $P_{mn}$ are quasi-periodic in $\vz$:
\[
P_{mn}(\vz+\vnu,\mu,\bp,\bq)=P_{mn}(\vz,\vmu,\bp,\bq)e^{-i\vnu\cdot\bq}.
\]
We use (\ref{p-ort}) and (\ref{delta-mu}) 
to obtain the orthogonality relation for the
functions $P_{mn}$:
\bea\label{orth-P} \sum_{\vmu\in
  L^*}\int_{R^d}\frac{d\vz}{|B|} P_{mn}(\vz,\vmu,\bp,\bq) \bar
P_{jl}(\vz,\vmu,\bp,\bq_0)=\delta_{mj}\delta_{nl}\delta_{per}(\bq-\bq_0).
\eea 
The operator ${\cal L}$ acts on these functions as
\be
{\cal L}P_{mn}=i(E_m(\bp)-E_n(\bp+\bq))P_{mn}(\vz,\vmu,\bp,\bq),
\ee
which is the analog of (\ref{l-on-qmn}) for the functions $Q_{mn}$. Note that
the integrand in (\ref{def-pmn}) is periodic in $\vy$ so no boundary terms are
produced by integration by parts.

We decompose $W_1$ in this basis as
\bea\label{w1-r}
W_1(t,\vx,\vz,\bp+\vmu)=\sum_{m,n}\int_{B}\frac{d\bq}{|B|}
\eta_{mn}(t,\vx,\bp,\bq)
P_{mn}(\vz,\vmu,\bp,\bq),
\eea
with $\vz\in R^d$, $\bp\in B$ and $\vmu\in L^*$.
We insert (\ref{w1-r})
into (\ref{order1/2-r}), multiply (\ref{order1/2-r}) by 
$\bar P_{jl}(\vz,\vmu,\bp,\bq_0)$, sum over $\vmu\in L^*$ and integrate
over $\vz\in R^d$. Then we get using (\ref{orth-P}) on the left side
\bea\label{eta-r}
\eta_{jl}(t,\vx,\bp,\bq_0)=
\sum_{\vmu\in L^*}\iint_{R^{2d}}
\frac{d\vz d\bq e^{i\bq\cdot\vz}\hat N(\bq)}{(2\pi)^{d}}
\frac{\left[W_0(\vz,\bp+\vmu-\bq)-
W_0(\vz,\bp+\vmu)\right]}{E_l(\bp+\bq_0)-E_j(\bp)+i\theta}
\bar P_{jl}(\vz,\vmu,\bp,\bq_0),~
\eea
where $\theta$ is the regularization parameter. We let $\theta\to 0$
at the end. We insert expression (\ref{decomp-3}) for $W_0$ and the definition
(\ref{qmn}) of $Q_m$ into (\ref{eta-r}). The resulting  expression may be 
simplified using (\ref{p-ort}) and (\ref{delta-mu}):
\bea
\eta_{jl}(t,\vx,\bp,\bq_0)&=&
\frac{1}{(2\pi)^d}\sum_{\vmu\in L^*}
\frac{\hat N(-\vmu-\bq_0)\sigma_l(\bp+\bq_0)}{E_l(\bp+\bq_0)-E_j(\bp)+i\theta}
A_{lj}(\bp+\bq_0+\vmu,\bp)\nonumber\\
\label{eta-r-3}
&-&\int_{R^d\times R^d}\frac{d\vz d\bq}{(2\pi)^d|B|}e^{i\bq\cdot\vz}
\frac{\hat N(\bq)\sigma_j(\bp)\Psi_l(\vz,\bp+\bq_0)\bar\Psi_j(\vz,\bp)}
{E_l(\bp+\bq_0)-E_j(\bp)+i\theta}.
\eea
Here the amplitude $A_{lj}(\bq,\bp)$ is given by
\bea\label{aij-r}
A_{lj}(\bq,\bp)=\int_C\frac{d\vy}{|C|}e^{-i(\bq-\bp)\cdot\vy}
\Psi_l(\vy,\bq)\bar\Psi_j(\vy,\bp).
\eea
The next order equation is 
\be
\pdr{W_0}{t}+\bk\cdot\nabla_{\vx}W_0+i\nabla_{\vz}\cdot\nabla_{\vx}W_0+
{\cal L}W_2=
\frac{1}{i}\int_{R^d}\frac{d\bq}{(2\pi)^d}e^{i\bq\cdot\vz}
\hat N(\bq)\left[W_1(\vz,\bk-{\bq})-W_1(\vz,\bk)
\right].
\ee
We multiply this equation by $\bar Q_{j}(\vz,\vmu,\bp)$, integrate over
$\vz$ and sum over $\vmu\in L^*$, 
and take average. Then as before the left side is
\bea\label{lhs-r}
\hbox{LHS}=\pdr{\sigma_j}{t}+
\nabla_\bp E_j\cdot\nabla_\vx\sigma_j.
\eea
The right side is
\bea\label{rhs-r}
\hbox{RHS}=I_1+I_2,
\eea
where
\bea\label{i1-1-r}
I_1=\frac 1i
\sum_{\vmu\in L^*}\int_C\frac{d\vz}{|C|}\int_{R^d}\frac{d\bq 
 e^{i\bq\cdot\vz}}{(2\pi)^d}\langle\hat N(\bq)
W_1(\bp+\vmu-{\bq})\bar Q_{j}(\vz,\vmu,\bp)\rangle
\eea
and
\be
I_2=-\frac 1i
\sum_{\vmu\in L^*}\int_C\frac{d\vz}{|C|}\int_{R^d}\frac{d\bq 
 e^{i\bq\cdot\vz}}{(2\pi)^d}\langle\hat N(\bq)
W_1(\bp+\vmu)\bar Q_{j}(\vz,\vmu,\bp)\rangle.
\ee
We insert expression (\ref{w1-r}) for $W_1$ into
(\ref{i1-1-r}) to get
\[
I_1
=\frac 1i
\sum_{\vmu\in L^*}\int_C\frac{d\vz}{|C|}\int_{R^d}\frac{d\bq}{(2\pi)^d}
\int_B\frac{d\bq_0}{|B|}\langle\hat N(\bq)\bar Q_j(\vz,\vmu,\bp)
\sum_{m,n}\eta_{mn}(\bp+\vmu-\bq,\bq_0)P_{mn}(\vz,\vmu,\bp-\bq,\bq_0)\rangle.
\]
We may split $I_1=I_{11}-I_{12}$ according to the two terms in (\ref{eta-r-3}).
We insert the first term in
(\ref{eta-r-3}) into the expression for $I_{11}$ and average 
using spatial homogeneity (\ref{covar-r}) of the random process $N(\vz)$, 
orthogonality (\ref{orth-psi}) of the Bloch functions 
$\Psi_m(\vz,\bp)$, and also sum over $\vmu\in L^*$ using (\ref{delta-mu}). 
Then we get
\be
I_{11}=\frac{1}{i}\sum_{\vmu\in L^*}\sum_m\int_B\frac{d\bq}{|B|(2\pi)^d}
\frac{\hat R(\bq+\vmu)|A_{jm}(\bp,\bp-\bq-\vmu)|^2\sigma_j(\bp)}
{E_j(\bp)-E_m(\bp-\bq)+i\theta}.
\ee
Here we have replaced integration over $\bq\in R^d$ by integration over 
$B$ and sum over $\vmu\in L^*$.
The second term $I_{12}$ is evaluated similarly:
\be
I_{12}=\frac{1}{i}\sum_{\vmu\in L^*}\sum_m\int_B\frac{d\bq}{|B|(2\pi)^d}
\frac{\hat R(\bq+\vmu)|A_{jm}(\bp,\bp-\bq-\vmu)|^2\sigma_m(\bp-\bq)}
{E_j(\bp)-E_m(\bp-\bq)+i\theta}.
\ee
Thus we have
\bea\label{i1-3-r}
I_1=\frac{1}{i}\sum_{\vmu\in L^*}\sum_m\int_B\frac{d\bq}{|B|(2\pi)^d}
\frac{\hat R(\bq+\vmu)|A_{jm}(\bp,\bp-\bq-\vmu)|^2[\sigma_j(\bp)-
\sigma_m(\bp-\bq)]}
{E_j(\bp)-E_m(\bp-\bq)+i\theta}.
\eea
One may verify that $I_2=\bar I_1$ in (\ref{rhs-r}). 
We insert then (\ref{i1-3-r}) into (\ref{rhs-r}), take the limit $\theta\to 0$,
make a change of variables $\bq\to\bp-\bq$,
and combine it with (\ref{lhs-r}) to get the system of radiative transport
equations:
\bea\label{rte-r}
\pdr{\sigma_j}{t}+\nabla_\bp E_j\cdot\nabla_\vx\sigma_j=
\sum_{m}
\int_B\frac{d\bq}
{|B|}{\cal Q}_{jm}(\bp,\bq)[\sigma_m(\bq)-
\sigma_j(\bp)]
\delta(E_j(\bp)-E_m(\bq)).
\eea
The differential scattering cross-sections ${\cal Q}_{jm}(\bp,\bq)$ are
given by
\[
{\cal Q}_{jm}(\bp,\bq)=\sum_{\vmu\in L^*}{1\over (2\pi)^{d-1}}
\hat R(\bp-\bq+\vmu)|A_{jm}(\bp+\vmu,\bq)|^2.
\]
This is the main result of this paper: we have derived a system of
coupled radiative transport equations for the phase space energy
densities of the Bloch modes. The transition probabilities 
${\cal Q}_{jm}(\bp,\bq)$ are real and symmetric: 
$\displaystyle {\cal Q}_{jm}(\bp,\bq)={\cal Q}_{mj}(\bq,\bp)$ as seen from
the definition (\ref{aij-r}) of $A_{ij}(\bp,\bq)$. 
Therefore the total energy is conserved:
\[ 
{\cal E}(t)=\sum_{m}\int_{R^d}d\vx\int_B d\bp
\sigma_m(t,\vx,\bp)=\hbox{const}.
\]
 Transport equations like (\ref{rte-r}) 
are well known in the theory of
resistance of metals and alloys \cite{Mott-Jones}.
Their systematic derivation from the Schr\"odinger equation
with a periodic and random potential (\ref{schro-rand}) is new.

\begin{appendix}
\section{Appendix. Multiple eigenvalues}

When eigenvalues $E_j(\bp)$ are not simple, but their multiplicity is
independent of $\bp$, so that there are still no level crossings, 
the analysis in the previous section can be
extended to this case.  This case is probably very rare for the Schr\"odinger 
equation but is important for other types of waves, for instance, in symmetric
hyperbolic systems. We present 
it here for the sake of completeness. The result is as follows. 
Let the matrices
$T_{jm}^{\al\beta}(\bp,\bq)$, $\al=1,\dots,r_j$, $\beta=1,\dots,r_m$,
where $r_j$ and $r_m$ are the multiplicities of $E_j$ and $E_m$, be
defined by \be T_{jm}^{\al\beta}(\bp,\bq)=\int_C \frac{d\vz}
{(2\pi)^{\frac{d-1}{2}}|C|}
e^{i(\bp-\bq)\cdot\vz}\Psi_m^\beta(\vz,\bq)\bar\Psi_j^\al(\vz,\bp).
\ee Then the coherence matrices $\sigma_j(\bp)$ satisfy the system of
radiative transport equations 
\bea
&&\pdr{\sigma_j}{t}+\nabla_\bp E_j\cdot\nabla_\vx\sigma_j\nonumber\\
&&= \sum_{\mu\in L^*\atop{m\in N}}\int_B\frac{d\bq}{|B|} \hat
R(\bp-\bq+\vmu)T_{jm}(\bp,\bq-\vmu)\sigma_m(\bq)
T_{jm}^*(\bp,\bq-\vmu)
\delta(E_j(\bp)-E_m(\bq))\nonumber\\
&&-\frac{i}{2\pi}
\int_B\frac{d\bq\hat R(\bp-\bq+\vmu)}{|B|}\left[
\frac{T_{jm}(\bp,\bq-\vmu)T_{jm}^*(\bp,\bq-\vmu)\sigma_j(\bp)}
{E_j(\bp)-E_m(\bq)+i0}\right.\left.-
\frac{\sigma_j(\bp)T_{jm}(\bp,\bq-\vmu)T_{jm}^*(\bp,\bq-\vmu)}
{E_j(\bp)-E_m(\bq)-i0}\right].\nonumber 
\eea 
These equations have
the same structure as the radiative transport equations for polarized
waves derived in \cite{RPK-WM}. The expression of the total scattering
cross-section as a principal value integral and not as a familiar
integral against $\delta[E_m(\bq)-E_j(\bp)]$ is known in transport theory 
for polarized waves
\cite{Howe-73,FR,RPK-WM}. They reduce typically to the form  common in
scalar transport equations under additional symmetries, like 
rotational invariance of the original wave equations and the power spectrum
tensor.

\end{appendix}

\section*{Acknowledgment}
The work of G.Bal and G.Papanicolaou
was supported by AFOSR grant F49620-98-1-0211 and by NSF grants DMS-9622854.
The work of A.Fannjiang was supported by NSF
grant DMS-9600119.
The work of L.Ryzhik 
was supported in part by the MRSEC Program
of the National Science Foundation under Award Number
DMR-9400379.


\begin{thebibliography}{99}

\bibitem{AS}
J. Avron and B. Simon, Analytic properties of band functions, 
Ann. Phys., {\bf 110}, 1979, 85-101.

\bibitem{BLP}
A. Bensoussan, J.-L. Lions and G. Papanicolaou, Asymptotic Analysis
of Periodic Structures, North Holland, 1978.


\bibitem{DA}
G. Dell'Antonio, Large time small coupling behavior of
a quantum particle in a random potential, Annals of Inst. H. Poincare,
section A, {\bf 39}, 1983, 339-384.



\bibitem{Erdos-Yau}
L. Erdos and H.T. Yau, Linear Boltzmann equation as scaling limit of 
quantum Lorentz gas, Preprint, 1997.

\bibitem{AF}
A. Fannjiang, unpublished notes, 1991.

\bibitem{FR}
A. Fannjiang and L. Ryzhik, Phase space transport theory for sound waves in 
random flows, Submitted to SIAM J. Appl. Math., 1998. 

\bibitem{G1}
P. G\'erard, Microlocal defect measures, Comm. PDEs, {\bf 16}, 1991, 1761-1794

\bibitem{G2}
P. G\'erard, Mesures Semi-Classiques et Ondes de Bloch, S\'em. Ecole 
Polytechnique, expos\'e {\bf XVI}, 1990-91, 1-19.


\bibitem{GL}
P.G\'erard, E.Leichtnam, Ergodic properties of eigenfunctions for the
Dirichlet problem, Duke Math. J., {\bf 71}, 1993, 559-607.

\bibitem{GMMP}
P. G\'erard, P. Markowich, N. Mauser and F. Poupaud, Homogenization limits and
 Wigner transforms, Comm.Pure Appl. Math., {\bf 50}, 1997, 323-380.



\bibitem{HLW}
T. Ho, L. Landau and A. Wilkins, On the weak coupling limit
for a Fermi gas in a random potential, Rev. Math. Phys., {\bf 5}, 1993,
209-298.
\bibitem{Howe-73}
M.Howe, Multiple scattering of sound by turbulence and other inhomogeneities,
Jour. Sound and Vibration {\bf 27}, 1973, 455-476.

\bibitem{Keller-Odeh}
J.B. Keller and F.Odeh, 
Partial differential equations with periodic coefficients and Bloch waves in
crystals. J. Math. Phys., 5, 1964, 1499--1504. 

\bibitem{KL}
J.B. Keller and R. Lewis, Asymptotic methods for
partial differential equations: The reduced wave equation and Maxwell's
equations, in Surveys in applied mathematics, eds. J.B. Keller,
D. McLaughlin and G. Papanicolaou, Plenum Press, New York, 1995.

\bibitem{MMP}
P. Markowich, N. Mauser and F. Poupaud, A Wigner function approach
to semiclassical limits: electrons in a periodic potential, Jour. Math. Phys.,
{\bf 35}, 1994, 1066-1094.

\bibitem{ME}
P. Martin and G. Emch, A rigorous model sustaining van Hove's
phenomenon, Helv. Phys. Acta, {\bf 48}, 1975, 59-78.

\bibitem{Mott-Jones}
N.F. Mott and H. Jones,
{\em The Theory of the Properties of Metals and Alloys},
Dover, NY, 1958.

\bibitem{CPDE}
L. Ryzhik, G. Papanicolaou and J.B. Keller,
Transport equations in a half space, Comm. PDE's, {\bf 22}, 1997, 1869-1911.


\bibitem{RPK-WM}
L. Ryzhik, G. Papanicolaou and J.B. Keller, 
Transport equations for elastic and 
other waves in random media, Wave Motion, {\bf 24}, 1996, 327-370.

\bibitem{Spohn}
H. Spohn, Derivation of the transport equation for electrons
moving through random impurities, Jour. Stat. Phys., {\bf 17}, 1977, 385-412.

	
\bibitem{Wigner}
E. Wigner, On the quantum correction for
thermodynamic
equilibrium, Physical Rev., {\bf 40}, 1932, 749-759.


\bibitem{WX}
C. Wilcox, Theory of Bloch waves, Journal d'Analyse Mathematique, {\bf 33},
1978, 146-167.

\bibitem{WX-2}
C. Wilcox, Scattering theory for diffraction gratings, Springer-Verlag, 
New York,1984. 

\end{thebibliography}
\end{document}